\documentclass[12pt,twoside,a4paper]{article}
\usepackage{amsmath,amssymb,latexsym,theorem,natbib,epsfig,color,subfigure,footmisc,bbm}
\usepackage{multirow,graphicx,array,rotating,url,wrapfig}  
\usepackage{epstopdf,afterpage,wasysym}
\usepackage{verbatim}
\usepackage[normalem]{ulem}
\def\rank{\mathop{\hbox{\rm rank}}}
\def\crps{\mathop{\hbox{\rm CRPS}}}

\def\logs{\mathop{\hbox{\rm LogS}}}

\def\es{\mathop{\hbox{\rm ES}}}

\def\vs{{\rm VS}}
\def\ri{\mathop{\hbox{\rm RI}}}

\numberwithin{equation}{section}

\title{Enhancing multivariate post-processed visibility predictions utilizing CAMS forecasts}

\author{{M\'aria Lakatos}$^{1,2}$  and {S\'andor Baran}$^{1,*}$ \vspace*{0.5cm}\\
{\small $^1$Faculty of Informatics, University of Debrecen, Hungary}\\
{\small $^2$Doctoral School of Informatics, University of Debrecen, Hungary}
}  

\date{}

\begin{document}

\maketitle

\footnotetext[1]{Corresponding author: \url{baran.sandor@inf.unideb.hu}}
\begin{abstract}

In our contemporary era, meteorological weather forecasts increasingly incorporate ensemble predictions of visibility -- a parameter of great importance in aviation, maritime navigation, and air quality assessment, with direct implications for public health. However, this weather variable falls short of the predictive accuracy achieved for other quantities issued by meteorological centers. Therefore, statistical post-processing is recommended to enhance the reliability and accuracy of predictions. By estimating the predictive distributions of the variables with the aid of historical observations and forecasts, one can achieve statistical consistency between true observations and ensemble predictions. Visibility observations, following the recommendation of the World Meteorological Organization, are typically reported in discrete values; hence, the predictive distribution of the weather quantity takes the form of a discrete parametric law. Recent studies demonstrated that the application of classification algorithms can successfully improve the skill of such discrete forecasts; however, a frequently emerging issue is that certain spatial and/or temporal dependencies could be lost between marginals. Based on visibility ensemble forecasts of the European Centre for Medium-Range Weather Forecasts for 30 locations in Central Europe, we investigate whether the inclusion of Copernicus Atmosphere Monitoring Service (CAMS) predictions of the same weather quantity as an additional covariate could enhance the skill of the post-processing methods and whether it contributes to the successful integration of spatial dependence between marginals. Our study confirms that post-processed forecasts are substantially superior to raw and climatological predictions, and the utilization of CAMS forecasts provides a further significant enhancement both in the univariate and multivariate setup.

\end{abstract}

\section{Introduction}
\label{sec1}
Due to its significant impact on a wide range of aspects of aviation meteorology, visibility is a critical parameter for both weather forecasters and pilots. Beyond the previously mentioned contexts, visibility's importance extends notably into the realms of agriculture and maritime transportation. Additionally, poor visibility conditions stand out as one of the most common causes of road accidents. The World Meteorological Organization (WMO) defines visibility as "the greatest distance at which a black object of suitable dimensions (located on the ground) can be seen and recognized when observed against the horizon sky" \citep{wmo92}. Due to the mentioned reasons, it is crucial to create the most accurate visibility forecasts possible, a task that, fortunately, an increasing number of meteorological services have been undertaking in recent years.

In the present day, weather forecasts are crafted through the use of numerical weather prediction (NWP) models. These models utilize initial values obtained from the previous states of the atmosphere, and the perturbation of these initial conditions results in the creation of a set of forecasts, commonly referred to as ensemble predictions. Despite notable advancements, NWP systems face challenges in achieving complete bias-free forecasts for various weather parameters, even within short lead times. Visibility emerges as a parameter with comparatively weak forecast skill, primarily because, unlike temperature, wind speed, or precipitation accumulation, most NWPs do not explicitly model visibility. Consequently, visibility forecasts necessitate derivation from predictions of related quantities such as relative humidity or precipitation \citep{cr11}. Specifically, at the European Centre of Medium-Range Weather Forecasts (ECMWF), the visibility parameter was integrated into the ECMWF Integrated Forecasting System (IFS) on 12 May 2015 \citep{ifs21}. This parameter utilizes model projections of water vapor, cloud, rain, and snow, along with climatological aerosol fields, to estimate the visibility that would be recorded by weather observers.

To correct the biases and lack of calibration in the forecasts, one possible direction is the application of statistical post-processing. In recent years, numerous post-processing techniques have been developed to calibrate ensemble forecasts, for a recent overview we refer to  \citet{vbd21}. From this wide range of methods, the straightforward yet potent parametric ensemble model output statistic \citep{grwg05} approach maps ensemble forecasts onto a probability distribution, providing a pathway for assessing uncertainties associated with potential outcomes. The commonly used Bayesian model averaging \citep[BMA;][]{rgbp05} also offers a complete predictive distribution for the examined weather variable; however, by now the machine learning-based distributional regression network \citep{rl18} method has became a rather popular parametric technique \citep[see e.g.][]{sl22}. Another widespread approach to post-processing is capturing the predictive distribution of a weather quantity with the help of its quantiles \citep[see e.g.][]{fh07,brem19}. Naturally, such nonparametric calibration is also frequently combined with machine learning techniques. As examples, one can mention the quantile regression forests \citep{tmzn16} or the Bernstein quantile network \citep{brem20}.

The utility of univariate post-processing techniques is supported by numerous studies; however, even today, many experts examine the disadvantage that the correlations between marginal distributions often get lost when applying such methods. This can be relevant to connections between locations, predictions initialized at the same time, or the loss of dependencies between weather quantities. Multivariate post-processing aims to restore potentially lost correlation structures, and there are already numerous parametric or non-parametric models available for achieving these goals \citep[see e.g.][]{sm18}. One option is to consider multivariate predictive distributions \citep[see e.g.][]{bm15}; however, the drawback of such models is that due to the estimation of a large number of parameters, they can easily encounter numerical problems. In contrast, the ensemble copula coupling \citep[ECC;][]{stg13} represents a two-step approach that involves initially generating samples from univariate predictive distributions. Subsequently, these samples are reorganized into the raw ensemble rank order structure, thus, this procedure can be interpreted as the application of an empirical copula. The Schaake Shuffle \citep{cghrw04} is another method that functions based on a similar principle to ECC; nevertheless, past research has suggested that ECC frequently demonstrates slightly better predictive performance \citep[see e.g.][]{cjsl24,llhb23}. For a comprehensive comparison of such multivariate post-processing methods, we refer to \citet{lbm20}.

Visibility observations are typically reported in discrete values according to WMO guidelines, namely "100 to 5,000 m in steps of 100 m, 6 to 30 km in steps of 1 km, and 35 to 70 km in steps of 5 km" \citep[Section 9.1.2]{wmo18}. As a result, the post-processing of visibility forecasts is reduced to a classification task, where the predicted probabilities of the different classes comprise the forecast distribution in the form of a discrete probability mass function (PMF). A recent study \citep{bl23} demonstrated that the implementation of multilayer perceptron neural network \citep[MLP;][]{dlbook} and proportional odds logistic regression \citep[POLR;][]{mccull80} classifiers can significantly enhance the forecast skill of raw ECMWF visibility predictions. Note, that these two methods displayed superior forecast skills also in the realm of post-processing ensemble predictions of total cloud cover \citep{bleab21}. However, in contrast to \citet{bleab21}, the investigation of \citet{bl23} did not include the exploration of the use of additional covariates that are not exclusively based on ECMWF forecasts of the target weather quantity. A natural first step in this direction is to consider further predictions of visibility that are independent of the operational ECMWF ensemble.

The Copernicus Atmosphere Monitoring Service (CAMS) is dedicated to monitoring and predicting the composition of Earth's atmosphere. It provides information on various atmospheric components, including air quality, greenhouse gases, and aerosols \citep{cams}. CAMS delivers deterministic forecasts related to these components, generated by extending the IFS model developed by ECMWF with additional modules. 

In this paper, we study the predictive accuracy of POLR and MLP techniques for calibrating ECMWF visibility ensemble forecasts, incorporating CAMS forecasts as covariates, in the contexts of both univariate and multivariate post-processing. Raw ECMWF visibility ensemble and climatology are considered as reference forecasts.

The structure of the paper is outlined as follows. In Section \ref{sec2}, a concise description of the visibility datasets under study is presented. Section \ref{sec3} provides a review of the considered univariate and multivariate post-processing methods, along with details on the approaches used for training data selection and the verification tools considered. The outcomes of our case study are detailed in Section \ref{sec4}, followed by a brief discussion and conclusions in Section \ref{sec5}.

\section{Data}
\label{sec2}

\begin{wrapfigure}{r}{0.5\textwidth}
  \begin{center}
  \vskip -1 truecm
    \epsfig{file=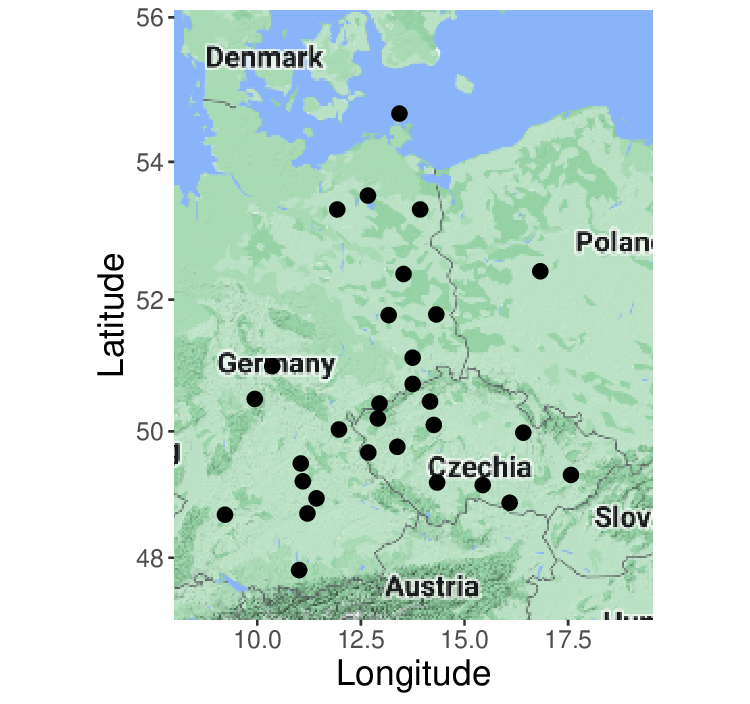, width=0.42\textwidth} 
   \label{fig:map}
  \end{center}
   \caption{Locations of SYNOP observation stations  corresponding to the ECMWF- and CAMS forecasts for 2020-2021}
\end{wrapfigure}

The visibility ensemble forecasts employed in this study consist of the operational ECMWF control (CTRL) and the corresponding 50 interchangeable ensemble members (ENS) produced by the ECMWF IFS. Regarding the latter 50 members of the ensemble, it is important to note that their interchangeability stems from the statistical indistinguishability arising from the perturbations of the initial conditions of the NWP. As mentioned in the Introduction, while visibility is primarily reported as a continuous variable in meters, the WMO suggests reporting observations in discrete categories, resulting in 84 different values in total, which can be described by the set
\begin{equation*}
\mathcal{Y} = \{0, 100, 200, \ldots, 4900, 5000, 6000, 7000, \ldots , 29000, 30000, 35000, 40000, \ldots , 65000, 70000\},
\end{equation*}
where the alignment of forecasts (provided in 1 m increments) with observations is achieved by rounding down to the nearest reported value. 

As discussed, our goal is to show that incorporating CAMS visibility forecasts as an additional predictor in the post-processing methods described in Section \ref{sec3}, -- representing an independent yet comparable meteorological parameter -- can significantly enhance the forecast skill of both the raw ECMWF predictions and the corresponding calibrated forecasts. The deterministic CAMS predictions are part of the global forecasts for atmospheric composition provided by the Copernicus Atmosphere Monitoring Service, a service provided by the ECMWF as part of the Copernicus program. The CAMS Global atmospheric composition forecast utilizes the IFS, the same model employed for ECMWF weather forecasts. However, specific modules within CAMS have been enabled to account for aerosols, reactive gases, and greenhouse gases, which have been developed uniquely within the CAMS framework.

Both the ECMWF ensemble, the CAMS forecasts, and the corresponding observations span a 2-year period from January 2020 to December 2021, for 30 SYNOP stations located in Germany, the Czech Republic, and Poland (refer to Figure \ref{fig:map}). All forecasts are initialized at 0000 UTC, and we consider 20 different lead times ranging from 6 hours to 120 hours, with a time step of 6 hours. Notably, this dataset is fairly complete, with no missing observations.

\section{Post-processing- and forecast evaluation methods}
\label{sec3}
In what follows, let \ $\boldsymbol f^{(d)} = \big(f_1^{(d)} , f_2^{(d)} ,\ldots , f_{51}^{(d)}\big)$ \ denote a 51-member ECMWF ensemble forecast for SYNOP station \ $d$  \ ($d = 1,2, \ldots, 30$) for a given time point with a given lead time, where \ $f_1^{(d)}=f^{(d)}_{\text{CTRL}}$ \ represents the control run, and \  $f^{(d)}_2,f^{(d)}_3, \ldots ,f^{(d)}_{51}$ \ correspond to the 50 exchangeable members \ $f^{(d)}_{\text{ENS},1},f^{(d)}_{\text{ENS},2}, \ldots ,f^{(d)}_{\text{ENS},50}$. \ Furthermore, let $f_0^{(d)}=f^{(d)}_{\text{CAMS}}$ be the matching CAMS prediction, while \ $Y^{(d)}\in{\mathcal Y}=\big\{y_1,y_2, \ldots ,y_{84}\big\}$ \ denotes the corresponding observed visibility with \ ${\mathcal Y}$ \ defined in Section \ref{sec2}. Finally, station-specific data are combined into multivariate forecasts \ $\boldsymbol{\mathfrak f}_0, \boldsymbol{\mathfrak f}_1, \ldots , \boldsymbol{\mathfrak f}_{51}$ \ and observations \ $\boldsymbol Y$, \ where \ $\boldsymbol{\mathfrak f}_{\ell} = \big(f_{\ell}^{(1)},f_{\ell}^{(2)}, \ldots , f_{\ell}^{(30)}\big)^{\top}, \ \ell =0,1, \ldots ,51,$ \ and \ $\boldsymbol Y =\big(Y^{(1)},Y^{(2)}, \ldots ,Y^{(30)}\big)^{\top}$.

\subsection{Univariate post-processing}
 \label{subs3.1}
To simplify the presentation of calibration methods for marginals, in this section we omit the indication of the SYNOP station and use notations \ $\boldsymbol f = \boldsymbol f^{(d)}$ \ and \ $Y = Y^{(d)}$ \ for forecasts and observations, respectively.

As \ $Y$ \ is a discrete random variable, the predictive distribution of \ $Y$ \ is specified by conditional probabilities
\begin{equation}
  \label{eq:predDist}
  {\mathsf P}(Y=y_k \mid \boldsymbol f), \qquad k=1,2,\ldots ,84,
\end{equation}
forming a conditional PMF with respect to the ensemble forecast.

Naturally, in \eqref{eq:predDist} the raw forecast can replaced by any feature vector \ $\boldsymbol x$ \ derived from the ensemble and/or other covariates. These covariates can be location or time-specific variables or additional forecasts of visibility or other related weather quantities.

\subsubsection{Multilayer perceptron neural network}
\label{subs3.1.1}
In recent years, the application of various types of artificial neural networks (ANNs) for the calibration of ensemble forecasts has become increasingly popular. These networks have a significant advantage compared to classical parametric or non-parametric approaches, as they are more flexible e.g. in representing non-linear relations between the ensemble forecasts and the parameters of the predictive distributions and in incorporating new explanatory variables in the modelling process. Multilayer perceptron (MLP) networks are a powerful class of ANNs widely employed for classification tasks in machine learning. A conventional MLP comprises multiple layers and nodes (neurons), where the value of each node is a result of the transformed (via an activation function) weighted sum of values from the nodes of the preceding layer, along with an added bias term. Input features are introduced to the input layer, and predictions regarding the distribution of different classes are generated in the output layer. The level of abstraction can be determined by tuning parameters, such as the number of hidden layers and the neurons within them. For more details, please refer to the work by \citet{dlbook}.

\subsubsection{Proportional odds logistic regression}
\label{subs3.1.2}
Proportional odds logistic regression (POLR) is an efficient parametric classification method for ordered classes, such as the visibility observations at hand. It specifies the conditional cumulative distribution of observed visibility \ $Y$ \ given an $M$-dimensional feature vector \ $\boldsymbol x$ \  as
\begin{equation*}
  {\mathsf P}\big(Y\leq y_k \mid \boldsymbol x\big)= \frac{{\mathrm e}^{\mathcal L_k(\boldsymbol x)}}{1+{\mathrm e}^{\mathcal L_k(\boldsymbol x)}},  \qquad \text{where} \qquad \mathcal L_k(\boldsymbol x):= \alpha_{k} + \boldsymbol x^{\top}\boldsymbol\beta, \quad \ k=1,2,\ldots, 84, 
\end{equation*}
with \ $\alpha_k\in {\mathbb R}, \ \boldsymbol \beta \in {\mathbb R}^M$, \ and \ $\alpha_{1}<\alpha_{2}< \cdots < \alpha_{84}$. \ Hence, POLR modelling of visibility classes results in a total of  \ $84+M$ \ unknown parameters to be estimated.

\subsection{Multivariate methods}
\label{subs3.2}
The purpose of applying univariate post-processing methods is to fine-tune and calibrate the forecasts predicted by the NWP model, addressing various accuracy issues or biases. However, when applying these techniques, there is a risk of losing the spatial, temporal, or even inter-variable correlations between marginals. Here we consider two easy-to-implement yet powerful two-step techniques that can correct this particular deficiency and restore dependence between marginal distributions \citep[see e.g.][]{lbm20, llhb23}. Both methods are based on reordering a sample generated from the calibrated predictive distribution obtained in the first step, where one can either consider equidistant quantiles or draw random or stratified samples. As \citet{cjsl24} reveal only minor differences in skill between the multivariate forecasts corresponding to the various sampling strategies, here we will work with the equidistant quantiles of the post-processed univariate forecasts.

\subsection{Ensemble copula coupling}
\label{subs3.2.1}
Ensemble copula coupling (ECC) leverages the raw ensemble rank structure (with ties resolved at random) and capitalizes on the ample information embedded in ensemble predictions concerning dependencies. Following the notations of \citet{llhb23} this iterative algorithm can be described as follows:

\begin{enumerate}
    \item Generate a sample \ $\widehat f_1^{(d)}, \widehat f_2^{(d)},  \ldots ,\widehat f_K^{(d)}$ \ for each dimension \ $d$, \ matching the size of the raw ensemble ($K$ = 51), from the calibrated marginal predictive distribution assumed to be sorted in ascending order.
    \item Consider permutations \ $\boldsymbol{\pi}_{d}=\big (\pi_{d}(1),\pi_{d}(2),\ldots ,\pi_{d}(K)\big)$ \ of \ $\{1,2,\ldots , K\}$ \ corresponding to the rank order structure of the raw ensemble \  $f_1^{(d)}, f_2^{(d)},  \ldots, f_K^{(d)}$, \ namely \ $\pi_{d}(k):=\rank \big(f_k^{(d)}\big)$, \ where ties are resolved at random. The ECC calibrated sample \ $\widetilde f_1^{(d)}, \widetilde f_2^{(d)},  \ldots ,\widetilde f_K^{(d)}$  \ for location \ $d$ \ is obtained by rearranging the sample generated in step~1 according to permutation  \ $\boldsymbol{\pi}_{d}$, \ that is 
  \begin{equation*}
    \widetilde f_k^{(d)}:=\widehat f_{\pi_{d}(k)}^{(d)}, \qquad k=1,2, \ldots, K, \quad d=1,2,\ldots, 30.
  \end{equation*}
\end{enumerate}  

\subsection{Schaake shuffle}
\label{subs3.2.2}
The other nonparametric multivariate post-processing technique we consider is the Schaake shuffle \citep[SSh;][]{cghrw04}, which involves using randomly selected past observations to create a dependence template aimed at restoring correlations between the marginal distributions. Samples from the calibrated univariate predictive distributions are then rearranged to match the rank order structure of this selected set of historical observations of the corresponding cardinality. In this way, the SSh method offers the flexibility to generate post-processed forecasts of various sizes, provided there is a sufficiently long historical climatological dataset available. However, to ensure a fair comparison with ECC methods, we restrict the sample size to that of the raw ensemble and use the same sampling strategy.

\subsection{Training data selection}
\label{subs3.3}
Each statistical post-processing method, including both the MLP and POLR models introduced in Section \ref{subs3.1}, conducts parameter estimation using training data, which comprises a collection of prior forecasts and observations predating the current forecast time. Various approaches exist for selecting this training dataset in terms of both spatial and temporal decomposition. 

A frequently used method of spatial selection, known as local estimation, entails using forecasts and observations specific to the station under consideration to estimate model parameters or weights \citep{tg10}. This approach could offer significant advantages as it effectively integrates the station's characteristics and local conditions into the calibration process. Another widely used and quite powerful alternative to the latter selection process is to treat together historical data of all stations of the ensemble domain, which, for a short training period, can be more efficient than local modelling. Nevertheless, this method, referred to as regional, is not particularly suitable for expansive and diverse domains, since it may struggle to effectively capture the varied complexities and nuances present across such landscapes. Due to the factors mentioned, in the case of regional modeling, all locations share the same set of parameters on a specific day to derive the parameters of the predictive distribution. A middle ground between regional and local methods is semi-local modeling, where stations with similar characteristics are clustered together, and regional modelling is applied within these clusters \citep{lb17}. 

The efficiency of post-processing methods is further influenced by the length of the training period and where the training period is positioned in time. One commonly used technique for temporal selection is when the training data is selected using a so-called sliding window approach, where the data for the preceding \ $n$ \ calendar days before the actual forecast date are chosen. Alternatively, fixing the training period to a specific extensive time frame in the past can also be suitable for implementing the modelling process, especially in the case of machine learning-based techniques requiring large training datasets. For a recent comparison of various time-adaptive training schemes, we refer to \citet{llmssz20}.

\subsection{Verification scores}
\label{subs3.4}
Following the suggestions of \citet{gr07}, the predictive performance of the studied univariate and multivariate visibility forecasts is evaluated with the help of proper scoring rules, as they simultaneously address both the calibration and sharpness of the predictive distribution. To be consistent with \citet{bl23}, in the univariate case, we consider the logarithmic score \citep[LogS;][]{good52} and the continuous ranked probability score \citep[CRPS;][Section 9.5.1]{w19}. For a discrete visibility forecast \ $F$ \ represented by a PMF \ $p_F(y)$ \ on \ $\mathcal Y$ \ the former is defined as 
\begin{equation*}
\logs(F,y) := -\log \big(p_F(y)\big),
\end{equation*}
whereas the latter equals
\begin{equation*}
\crps(F,y) = \sum_{k=1}^{84}p_F(y_k)\big| y- y_k\big| - \sum_{k=2}^{84}\sum_{\ell =1}^{k-1}p_F(y_k)p_F(y_\ell)\big| y_{\ell}- y_k\big|,
\end{equation*}
which is the discrete version of the general representation
\begin{equation}
\label{eq:crpsGen}
\crps(F,y) = {\mathsf E}|Y - y| -\frac 12{\mathsf E} |Y-Y'|
\end{equation}
derived by \citet{gr07}, where \ $Y$ \ and \ $Y'$ \ are independent random variables with finite second moments distributed according to \ $F$.

Form \eqref{eq:crpsGen} allows a direct extension of the CRPS to multivariate forecasts. For a $D$-dimensional predictive cumulative distribution function (CDF) \ $F$ \ and vector \ $\boldsymbol{y}=\big(y^{(1)},y^{(2)},\ldots ,y^{(D)}\big)^{\top},$ \ the energy score \citep[ES;][]{gr07} is defined as
\begin{equation*}
\es(F,\boldsymbol y)={\mathsf E}\Vert \boldsymbol Y - \boldsymbol y\Vert -\frac 12{\mathsf E} \Vert \boldsymbol Y-\boldsymbol Y'\Vert,
\end{equation*}
where now \ $\boldsymbol Y$ \ and \ $\boldsymbol Y'$ \ are independent random vectors having distribution \ $F$, and \ $\Vert \cdot \Vert$ denotes the Euclidean distance. For a \ $K$-member forecast ensemble or sample drawn from the predictive distribution, one has to consider the empirical CDF \ $F_K$ \ resulting in the ensemble energy score \citep{gsghj08}
\begin{equation*}
\es(F_K,\boldsymbol y)=\frac 1K \sum _{j=1}^K\Vert \boldsymbol
{\mathfrak f}_j-\boldsymbol y\Vert-\frac 1{2K^2}\sum_{j=1}^K\sum_{k=1}^K\Vert
\boldsymbol {\mathfrak f}_j-\boldsymbol {\mathfrak f}_k\Vert.
\end{equation*}

Finally, we also report the (ensemble) variogram score of order \ $p$ \citep[$\vs^p$;][]{sh15}, which is more sensitive to the errors in the specification of correlations than the $\es$.  Given an ensemble forecast \  $\boldsymbol {\mathfrak f}_k=\big(f_k^{(1)},f_k^{(2)}, \ldots ,f_k^{(D)}\big)^{\top}, \ k=1,2, \ldots ,K,$  \ it equals
\begin{equation*}
  \vs^p(F_K,\boldsymbol y)=\sum_{i=1}^D\sum_{j=1}^D \omega_{ij}\left(\big| y^{(i)}-y^{(j)}\big|^p  - \frac 1K\sum_{k=1}^K \big| f_k^{(i)}-f_k^{(j)}\big|^p\right)^2,
\end{equation*}
where \ $\omega_{ij}\geq 0$ \ is the weight for coordinate pair \ $(i,j)$. \ The usual choices for order \ $p$ \ are $0.5$ and $1$, in our case study we consider the former and use the notation \ $\vs$ \ for \ $\vs^{0.5}$. \ Note that all four proper scoring rules defined above are negatively oriented, that is the smaller the better, and implemented in the {\tt R} package {\tt scoringRules} \citep{jkl19}.

In the case study of Section \ref{sec4} the predictive performance of a forecast \ $F$ \ with a given forecast horizon in terms of a score \ $\mathcal S_F$ \ is quantified by the mean score value \ $\overline{\mathcal S}_F$ \ over all forecast cases used for verification. Furthermore, we also consider skill scores \citep{murphy73}
\begin{equation*}
     {\mathcal {SS}}_F :=1-\frac{\overline{\mathcal S}_F}{\overline{\mathcal S}_{F_\text{ref}}}, 
\end{equation*}
providing the relative improvement of a forecast \ $F$ \ in terms of the score \ $\mathcal S_F$ \ with respect to a reference forecast \  \ $F_\text{ref}$ \ resulting in a mean score value  \ $\overline{\mathcal S}_{F_\text{ref}}$. \ Skill scores are positively oriented (the larger the better) and often reported in percentage values. In Section \ref{sec4} we investigate continuous ranked probability skill score (CRPSS), logarithmic skill score (LogSS), energy skill score (ESS), and variogram skill score (VSS).

In addition to reporting the various score and skill score values of the competing visibility forecasts, we also address the significance of the score differences and uncertainty in score values by providing 95\,\% confidence bounds. These confidence intervals are calculated using 2000 block-bootstrap samples according to the stationary bootstrap scheme with the mean block length formula specified by \cite{pr94}.

Calibration of univariate ensemble forecasts can also be assessed graphically with the help of verification rank histograms \citep[][Section 9.7.1]{w19}. For a properly calibrated $K$-member ensemble, the ranks of the verifying observations with respect to the corresponding forecasts follow a discrete uniform distribution on the set \ $\{1,2, \ldots ,K+1\}$, \ and the shape of the corresponding histogram reflects the source of the lack of calibration. Multivariate generalizations of the verification rank are based on the concept of pre-ranks and the corresponding histograms display the pre-rank of the observation with respect to the ensemble forecast with ties resolved at random \citep{gsghj08}. Here we consider the average and band-depth ranks suggested by \citet{tsh16} and energy score and dependence (variogram) ranks studied in detail e.g. by \citet{azg24}. The average rank is simply the mean of the univariate ranks and the interpretation of the corresponding histogram is similar to the univariate case: $\cup$- and $\cap$-shaped histograms refer to under- and overdispersion, respectively, whereas biased forecasts result in triangular shapes. Band-depth histograms are based on the discrete version of ordering multivariate functional data proposed by \citet{lpr09}, while energy score and dependence histograms use the idea of \citet{kkp22} of proper score-based pre-ranks. Right-skewed band-depth histograms indicate overdispersion, left-skewed ones underdispersion or bias, whereas $\cup$- and $\cap$-shaped histograms mean too low- and high correlations in the ensemble. Furthermore, low score-based ranks indicate that the observation is similar to the ensemble members in terms of the given scoring rule, whereas outliers result in high pre-ranks. Note, that in the case of the dependence ranking the weight for a pair of coordinates in the variogram score is the negative exponential of the geographical distance between the corresponding SYNOP stations given in 100 km. Finally, as a measure of the deviation of verification ranks from uniformity we consider the reliability index \citep[RI;][]{dmhzds06} defined as
\begin{equation}
   \label{eq:relind}
 \ri:=\sum_{r=1}^{K+1}\Big| \rho_r-\frac 1{K+1}\Big|,
\end{equation}
where \ $\rho_r$ \ is the relative frequency of rank \ $r$.

\section{Results}
\label{sec4}
We first explore the advantage of applying CAMS visibility forecasts as additional covariates in POLR and MLP methods for discrete post-processing of ECMWF ensemble predictions of visibility compared to models based on the raw ECMWF ensemble only. We also study the predictive performance of the corresponding multivariate forecasts obtained with the help of ensemble copula coupling and Schaake shuffle. To be consistent with the results of \citet{bl23}, all investigated MLP and POLR classifiers rely on 350-day rolling training periods, and each lead time is treated separately. We consider local (MLP-L and POLR-L), regional (MLP-R and POLR-R) and clustering-based semi-local (MLP-C and POLR-C) training with four clusters derived using $k$-means clustering of SYNOP stations with respect to three-dimensional feature vectors consisting of observed frequencies of visibility intervals 0--5000, 5000--30000, and 30000--70000 m in the training period. Both univariate and multivariate predictions are validated on forecast-observation pairs for calendar year 2021 and as reference forecasts, we consider the raw ECMWF visibility ensemble and 30-day climatology.

\begin{figure}[t]
\begin{center}
\epsfig{file=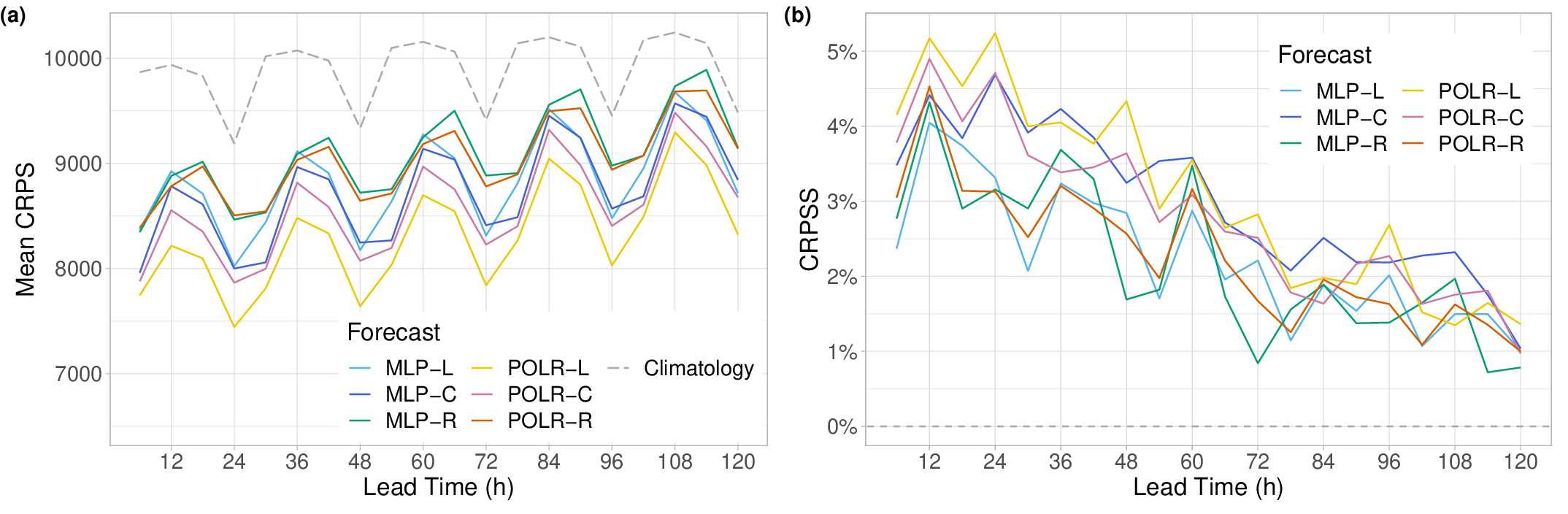, width=\textwidth}
\end{center}
\caption{Mean CRPS of climatological and post-processed visibility forecasts based on CAMS extended MLP and POLR models (a) and CRPSS of CAMS extended post-processed forecasts with respect to the corresponding MLP and POLR models based only on ECMWF predictions (b) as functions of the lead time.}
\label{fig:crps_crpss}
\end{figure}

The reference MLP and POLR models studied by \citet{bl23} are based on eight-dimensional input feature vectors 
\begin{equation}
\label{eq:feat}
\big(\tilde f_{CTRL},\overline f_{ENS},s^2,p_1,p_2,p_3,\beta_1,\beta_2\big)^{\top},
\end{equation}
where  \ $\tilde f_{CTRL}:=f_{CTRL}/70000$ \ is the normalized control member,  \ $\overline f_{ENS}$ \ is the mean of the 50 (normalized) exchangeable members, \ $s^2$ \ is the variance of the 51-member (normalized) operational ensemble,  \ $p_1, \ p_2$ \ and \ $p_3$ \ are the proportions of ensemble members predicting visibility up to 1000 m, 1000--2000 m and more than 30000 m, respectively, whereas \ $\beta_1$ \ and \ $\beta_2$ \ are annual base functions addressing seasonal variations defined as
\begin{equation*}
    \beta_1(d):=\sin \big(2\pi d/365 \big) \qquad \text{and} \qquad  \beta_2(d):=\cos \big(2\pi d/365 \big),
  \end{equation*}
where \ $d$ \ denotes the day of the year \citep[see e.g.][]{dmmz17}.

Here we add the normalized CAMS forecast \ $\tilde f_{CAMS}:=f_{CAMS}/70000$ \ to the input features and investigate the forecast skill of CAMS extended local, regional, and semi-local MLP and POLR post-processing based on the feature vector 
\begin{equation}
\label{eq:featEx}
\big(\tilde f_{CTRL},\overline f_{ENS}, \tilde f_{CAMS},s^2,p_1,p_2,p_3,\beta_1,\beta_2\big)^{\top}
\end{equation}
together with the corresponding multivariate forecasts derived with the help of the two-step approaches presented in Section \ref{subs3.2}.

\subsection{Univariate performance}
\label{subs4.1}

\begin{figure}[t]
\begin{center}
\epsfig{file=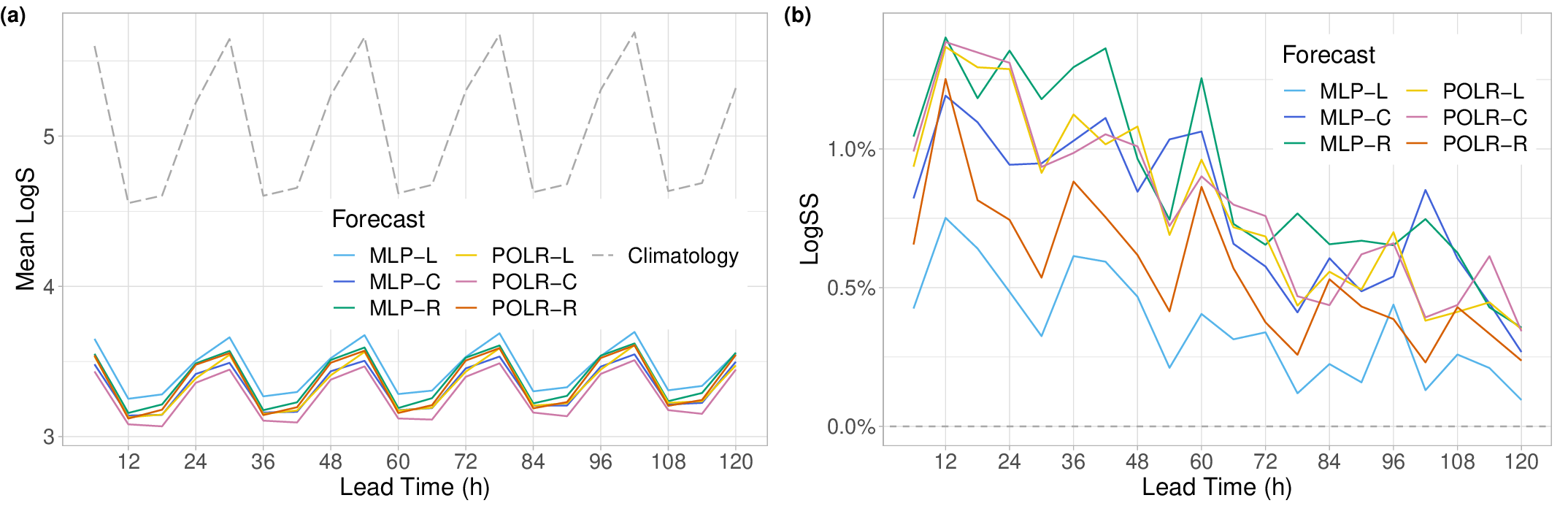, width=\textwidth}
\end{center}
\caption{Mean LogS of climatological and post-processed visibility forecasts based on CAMS extended MLP and POLR models (a) and LogSS of CAMS extended post-processed forecasts with respect to the corresponding MLP and POLR models based only on ECMWF predictions (b) as functions of the lead time.}
\label{fig:logs_logss}
\end{figure}

The predictive performance of univariate visibility forecasts is evaluated with the help of the mean CRPS and mean LogS over all forecast cases in the verification period, together with the corresponding skill scores. As in \citet{bl23}, extremely low predicted probabilities resulting in numerical issues in LogS calculations are handled by introducing a probability threshold \ $p_{\min} = 2.75\times 10^{-5}$, \ which replaces all values in the predictive PMFs which are below this probability. This particular choice of \ $p_{\min}$ \ ensures that with a 1\,\% probability the corresponding reported visibility materializes at least once in a year; however, it is low enough, so there are no observable differences in terms of other scores between the original and the adjusted and renormalized PMFs. Furthermore, as for the reference POLR approaches based on feature vector \eqref{eq:feat} the weights of \ $\tilde f_{CTRL}$ and \ $\overline f_{ENS}$ \ are forced to be non-negative, we apply the same constraint also for the coefficient of  \ $\tilde f_{CAMS}$ \ of the CAMS extended local, semi-local and regional POLR models. For further details of the implementation of MLP and POLR classifiers see \citet{bl23} and \citet{hhp16}.

\begin{figure}[t]
\begin{center}
\epsfig{file=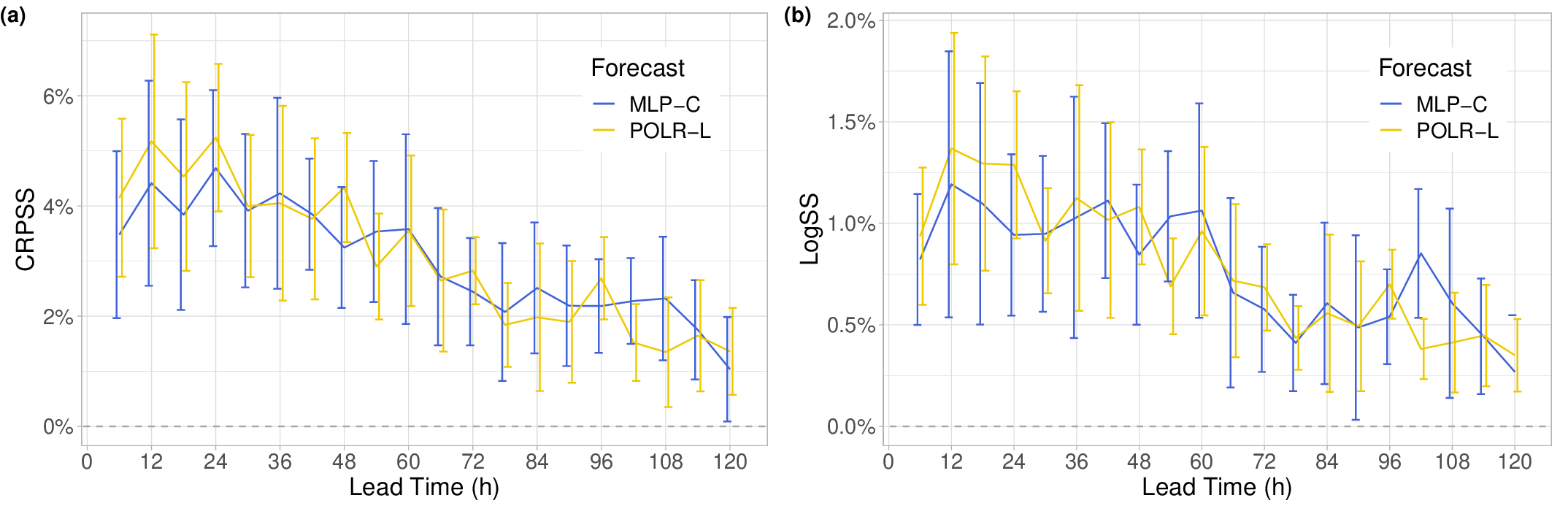, width=\textwidth}
\end{center}
\caption{CRPSS (a) and LogSS (b) of CAMS extended MLP-C and POLR-L forecasts with respect to the corresponding models based only on ECMWF predictions together with 95\,\% confidence bounds as functions of the lead time.}
\label{fig:crpss_logss_EB}
\end{figure}

As demonstrated in Section 4.2 of \citet{bl23}, in terms of the mean CRPS and mean LogS, up to 120 h, all MLP and POLR post-processed forecasts have shown significant improvement compared to both the raw ensemble and climatology. The extent of this improvement is particularly striking in the case of the raw ensemble, which we have chosen to exclude from our univariate analysis. Figure \ref{fig:crps_crpss}a displays the mean CRPS of climatological and CAMS extended MLP and POLR forecasts as functions of the forecast horizon. All post-processed predictions outperform climatology for all lead times, and the ranking of the various approaches is identical to the one for the corresponding reference models based on feature vector \eqref{eq:feat}, see Figure 2 of \citet{bl23}. The advantage achieved in mean CRPS by utilizing the CAMS extended feature vector \eqref{eq:featEx} over models based solely on ECMWF forecasts is illustrated by the CRPSS values of Figure \ref{fig:crps_crpss}b. In general, it can be observed that the integration of CAMS into the post-processing workflow for regional models yields the lowest benefits, but this advantage is still noteworthy. Note that the best performing POLR and MLP variants, which are the local POLR model and the semi-local MLP, profit the most from the extension of the input features with an average CRPSS of around 3\,\%. One should also remark that the improvement in mean CRPS from including CAMS predictions might further decrease for forecast horizons beyond 120 h. 

The mean LogS values of Figure \ref{fig:logs_logss}a tell the same story about the various post-processing approaches as Figure 3 of \citet{bl23}. Adding the CAMS forecast to the input features of the local, semi-local, and regional MLP and POLR models does not change the ranking of the forecasts; however, as portrayed in Figure \ref{fig:logs_logss}b, it results in a modest 0.4 -- 0.9\,\% average improvement in LogS. 

To address the significance of gain of CAMS extended post-processing models with respect to the corresponding forecasts based only on ECMWF predictions, in Figure \ref{fig:crpss_logss_EB} we accompany the CRPSS and LogSS of the MLP-C and POLR-L approaches, which are the best model variants according to \citet{bl23}, by 95\,\% block-bootstrap confidence intervals. As depicted in Figures \ref{fig:crpss_logss_EB}a and \ref{fig:crpss_logss_EB}b, up to the maximal studied lead time of 120 h, the improvement in terms of both scores is significant at a 5\,\% level.

\begin{figure}[t]
\begin{center}
\epsfig{file=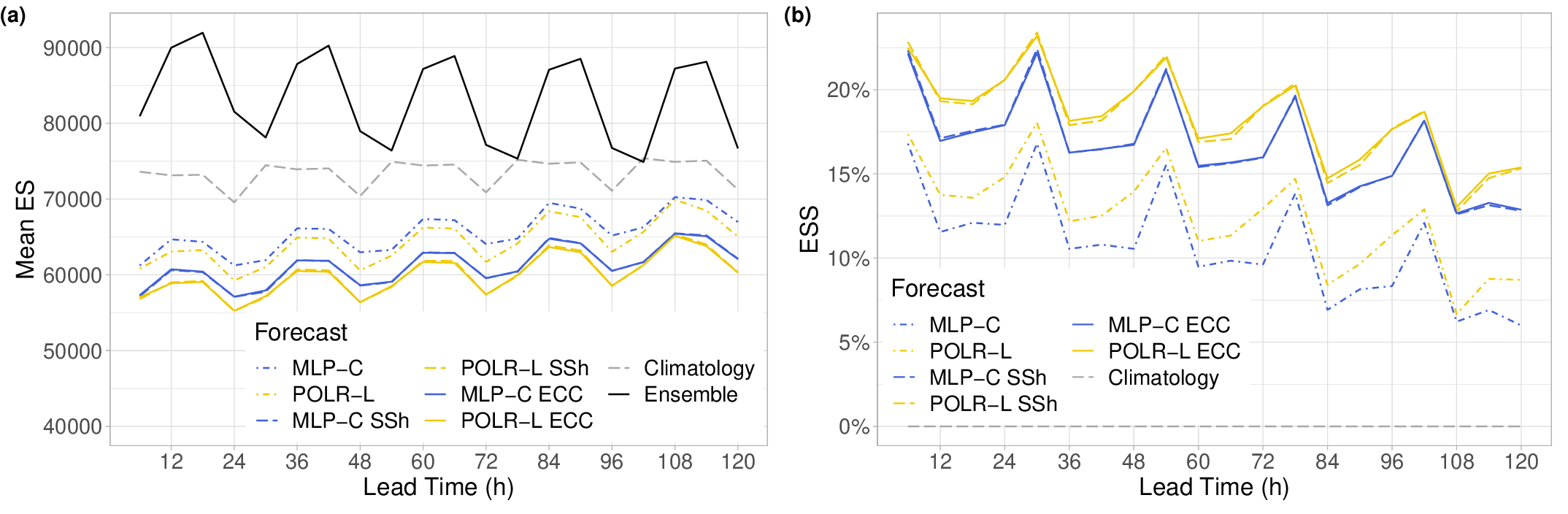, width=\textwidth}
\end{center}
\caption{Mean ES of raw, climatological and post-processed multivariate visibility forecasts based on CAMS extended MLP-C and POLR-L models (a) and ESS of CAMS extended post-processed forecasts with respect to multivariate climatology (b) as functions of the lead time.}
\label{fig:es_ess_clim}
\end{figure}

\subsection{Multivariate post-processing}
\label{subs4.2}
Continuing our analysis, we delve into the comparison of forecast skill among multivariate approaches detailed in Section \ref{subs3.2}. In the case of the applied multivariate methods, calibrated samples of size $51$ are employed, mirroring the dimensionality of the raw ECMWF ensemble. When employing the SSh method, which relies on historical data to establish dependency templates, we utilize observations obtained from the rolling training window applied during univariate calibration. This decision is made based on the finding of \cite{llhb23}, that extending these pool dates to encompass the entire available historical dataset may only yield negligible advantages. In our case, this entails selecting 51 days from a pool of 350 dates (which is the length of the training period). As mentioned, in both the ECC and SSh methods, the initial step entails generating samples from the calibrated PMFs, which we perform by considering their equidistant quantiles.

\begin{figure}[t]
\begin{center}
\epsfig{file=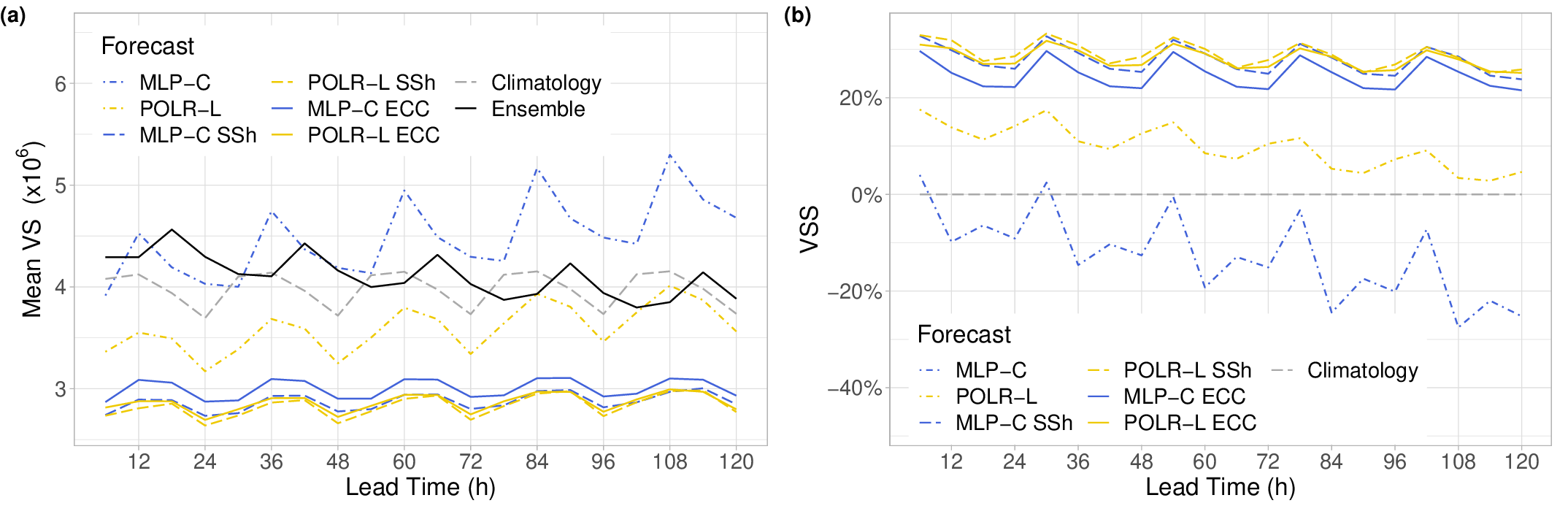, width=\textwidth}
\end{center}
\caption{Mean VS of raw, climatological and post-processed multivariate visibility forecasts based on CAMS extended MLP-C and POLR-L models (a) and VSS of CAMS extended post-processed forecasts with respect to climatology (b) as functions of the lead time.}
\label{fig:vs_vss_clim}
\end{figure}

To simplify presentation, as univariate calibrated forecasts we consider the CAMS extended MLP-C and POLR-L predictions and their counterparts based only on the ECMWF ensemble. The corresponding multivariate predictions obtained with the help of the ECC and SSh are referred to as MLP-C ECC, POLR-L ECC, and MLP-C SSh, POLR-L SSh, respectively. We also report the performance of naive multivariate forecasts obtained by simply arranging the 51-member station-specific MLP-C and POLR-L samples into 30-dimensional vectors. In the multivariate context, notations MLP-C and POLR-L will refer to these "independent" predictions. Finally, as a multivariate climatological forecast, we consider the vector of 51 equidistant quantiles of 30-day climatological PMFs corresponding to the investigated SYNOP stations. In principle, one could also consider the vectors of station-specific 51-day climatological forecasts; however, due to their discrete nature, they would substantially differ from the competing calibrated predictions. 

First, we assess the performance of the previously discussed multivariate forecasts based on the energy score. Figure \ref{fig:es_ess_clim}a clearly illustrates the substantial superiority of each CAMS extended multivariate forecast over the raw ensemble and the 30-day climatology in terms of this scoring rule. Notably, the semi-local MLP ECC and local POLR ECC exhibit the best predictive performance, extremely closely approaching their SSh counterparts, thereby minimizing performance disparities and highlighting the success of multivariate post-processing. According to the energy skill scores of Figure \ref{fig:es_ess_clim}b, the univariate MLP-C and POLR-L models alone demonstrate a substantial average improvement of 12\,\% over multivariate climatology, while 
the multivariate models show even more pronounced performance enhancements, averaging 17\,\%. This analysis emphasizes that, at least in our dataset, irrespective of the selected dependency template, multivariate models consistently exhibit improvement compared to independent calibration and climatology.

\begin{figure}[t]
\begin{center}
\epsfig{file=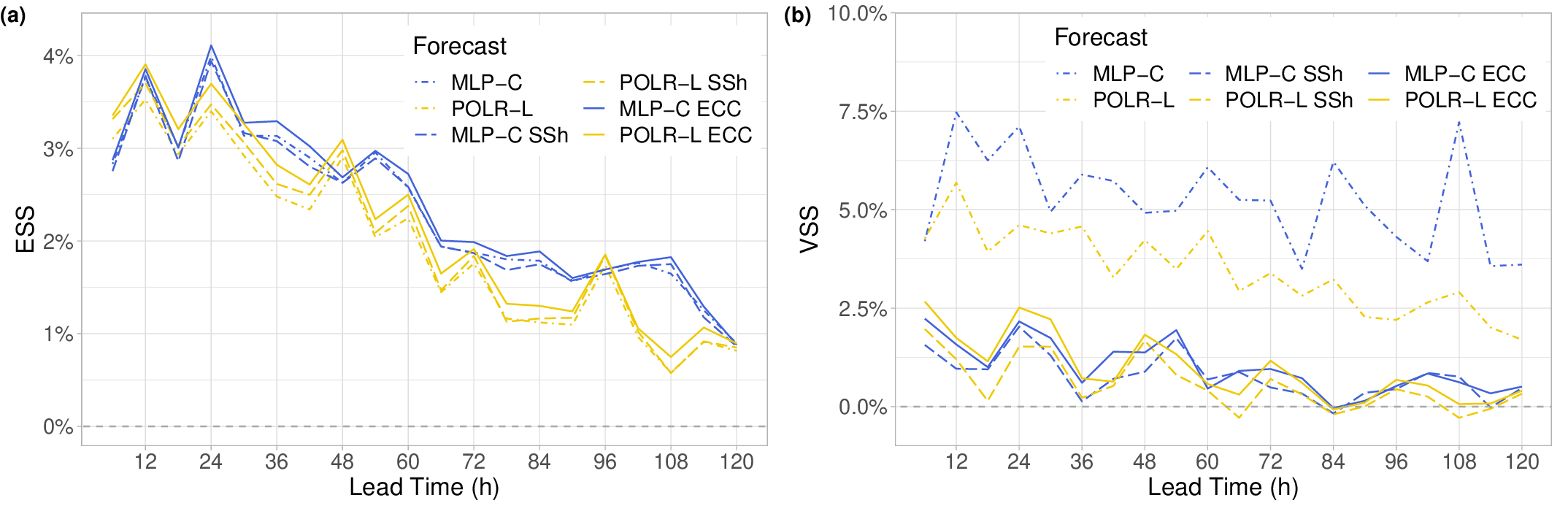, width=\textwidth}
\end{center}
\caption{ESS (a) and VSS (b) post-processed multivariate visibility forecasts based on CAMS extended MLP-C and POLR-L models with respect to the corresponding forecasts based only on ECMWF predictions as functions of the lead time.}
\label{fig:ess_vss}
\end{figure}

Figure \ref{fig:vs_vss_clim} presents a slightly different picture in terms of model ranking,
which might be a natural phenomenon when considering the variogram score depicted here. This verification measure clearly illustrates the success of multivariate post-processing in the redefinition of covariance structures. As expected, (CAMS extended) naive models, such as MLP-C and POLR-L, do not capture the spatial dependencies and might underperform the raw forecast vectors and multivariate climatology, as well. According to the VSS values of Figure \ref{fig:vs_vss_clim}b, the two-step
multivariate models demonstrate a notable improvement of over 20\,\% with respect to the reference climatological predictions. Among these approaches, the POLR variants exhibit the highest skill, particularly the POLR-L SSh configuration. Moreover, it is noteworthy that a more pronounced advantage was attained with the SSh models obtaining dependence templates from past observations, compared to the ECC forecasts utilizing the rank structure of raw ensemble.

In Figure \ref{fig:ess_vss}, we illustrate how the CAMS extended post-processed forecasts compare to the corresponding multivariate predictions from classifiers that ignore CAMS forecasts. Both the ESS and VSS reveal a diminishing advantage of CAMS integration with increasing forecast horizons. In the case of the ES the differences between the various post-processed forecasts are minor (see Figure \ref{fig:ess_vss}a) with the MLP-C ECC variant showing the most overall improvement. The VSS values of Figure \ref{fig:ess_vss}b again tell a different story. With independent MLP-C and POLR-L models, there is an average improvement of around 5\,\% in variogram scores suggesting that using CAMS as a feature helps in better preserving spatial dependencies. For the multivariate models, the gain is much smaller, especially for the 84-hour horizon and the POLR-L SSh model even displays negative skill at 66 and 108 hours.

\begin{figure}[t]
\begin{center}
\epsfig{file=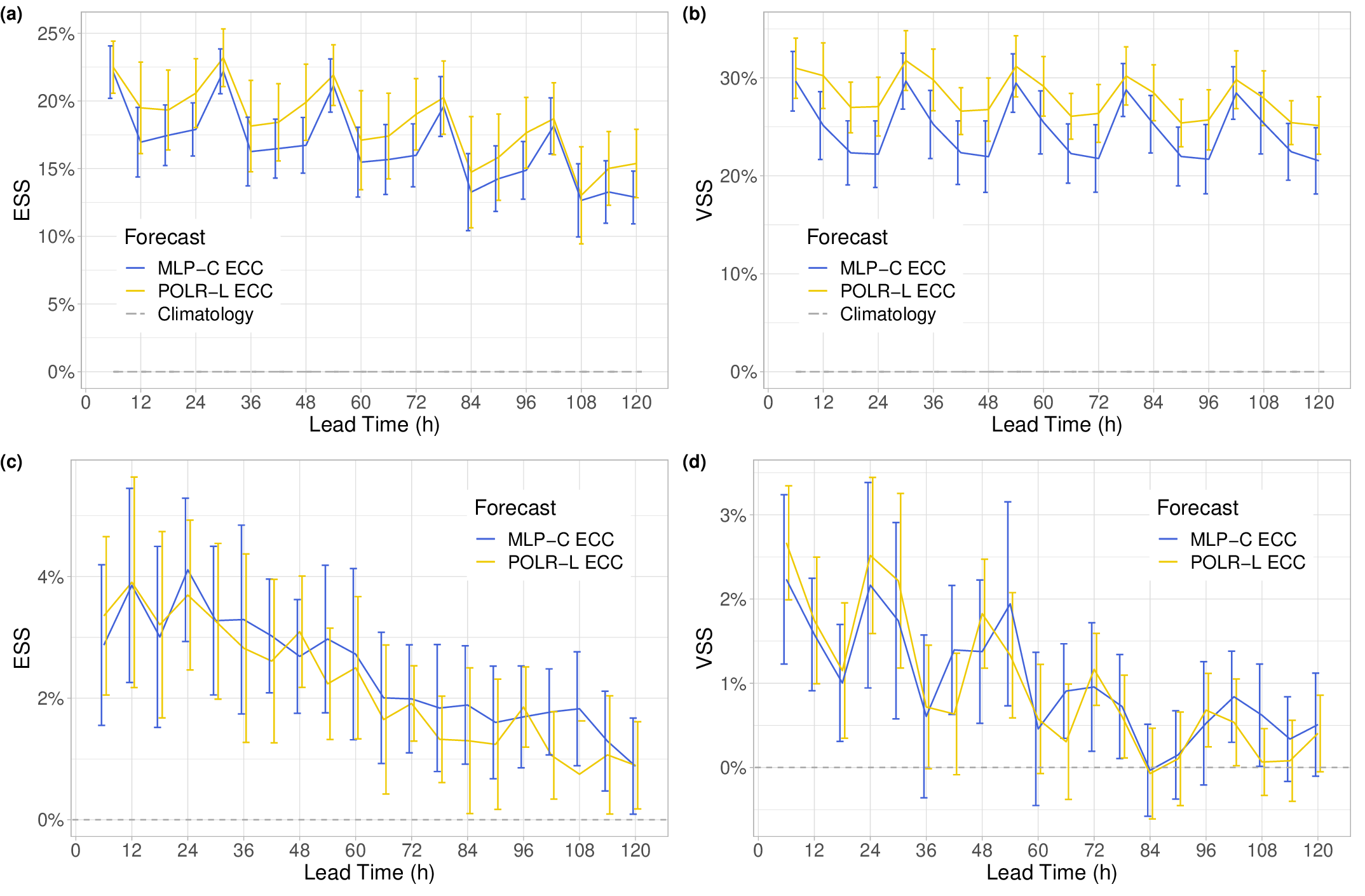, width=\textwidth}
\end{center}
\caption{ESS (a,c) and VSS (b,d) of CAMS extended MLP-C ECC and POLR-L ECC multivariate forecasts with respect to multivariate climatology (a,b) and to the corresponding MLP-C ECC and POLR-L ECC forecasts based only on ECMWF predictions (c,d) together with 95\,\% confidence bounds as functions of the lead time.}
\label{fig:ess_vss_EB}
\end{figure}

For better clarity, in Figure \ref{fig:ess_vss_EB}, we accompanied the ESS and VSS values of MLP-C ECC and POLR-L ECC forecasts -- which are overall considered the best performers -- with 95\,\% bootstrap confidence bounds. In cases where climatology serves as the reference, as seen in Figures \ref{fig:ess_vss_EB}a and \ref{fig:ess_vss_EB}b, in general, the local POLR-based ECC outperforms its MLP-C counterpart. However, for forecasts corresponding to 6 UTC, both in terms of ES and VS, the difference is not significant, indicating that the advantage of the POLR-L ECC approach over the MLP-C ECC is more apparent for the other three studied observation times (12, 18, 24 UTC). Furthermore, Figures \ref{fig:ess_vss_EB}c and \ref{fig:ess_vss_EB}d demonstrate that incorporating CAMS forecasts as an additional feature leads to a significant enhancement in ES for all investigated lead times, whereas for VS there are a few instances of longer lead times exhibiting negative skill within the confidence bounds.

\begin{figure}[h!]
\begin{center}
\epsfig{file=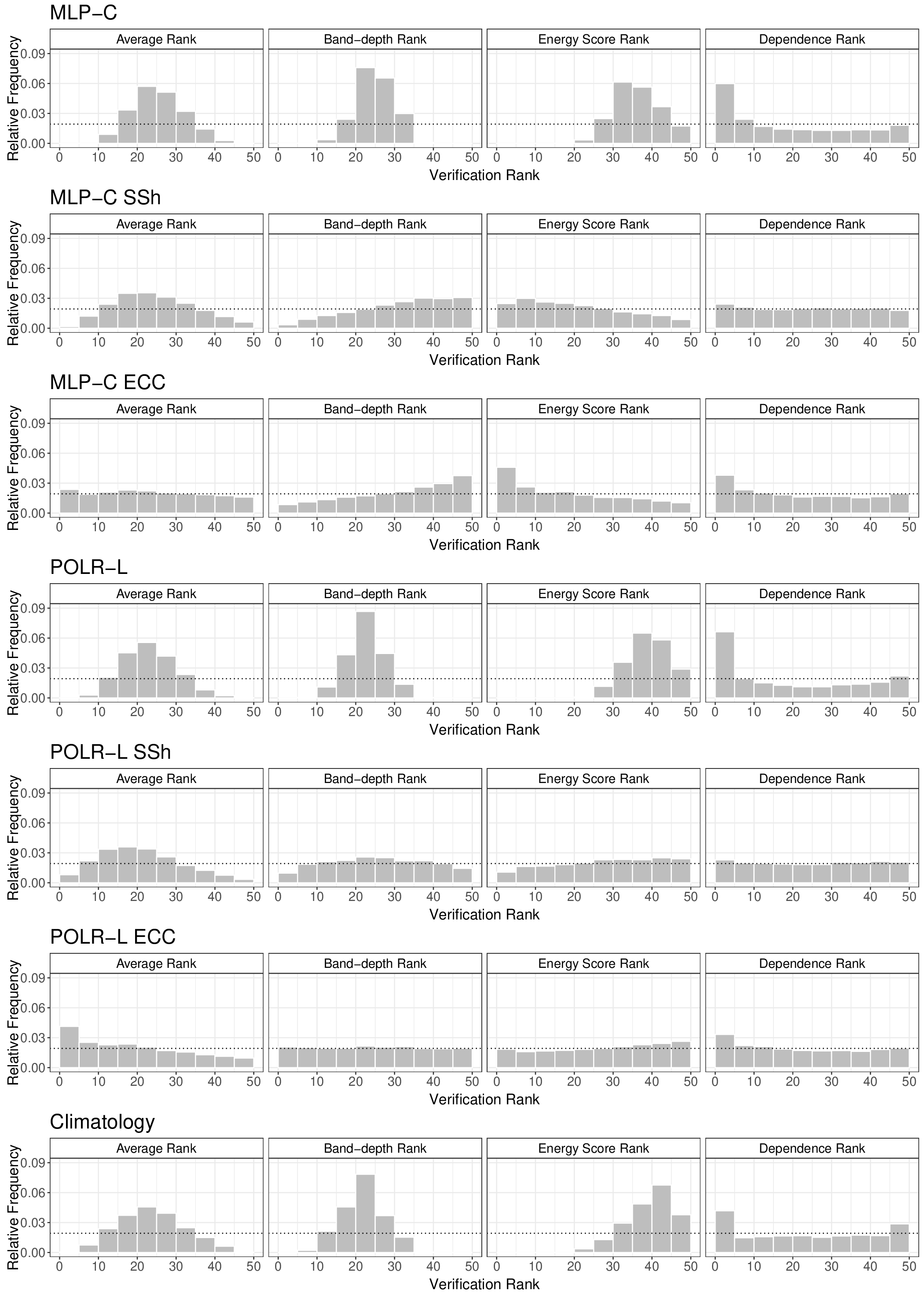, width=.92\textwidth}
\end{center}
\caption{Rank histograms of climatological and post-processed multivariate visibility forecasts based on CAMS extended MLP-C and POLR-L models.}
\label{fig:rankhist}
\end{figure}

\begin{figure}[t]
\begin{center}
\epsfig{file=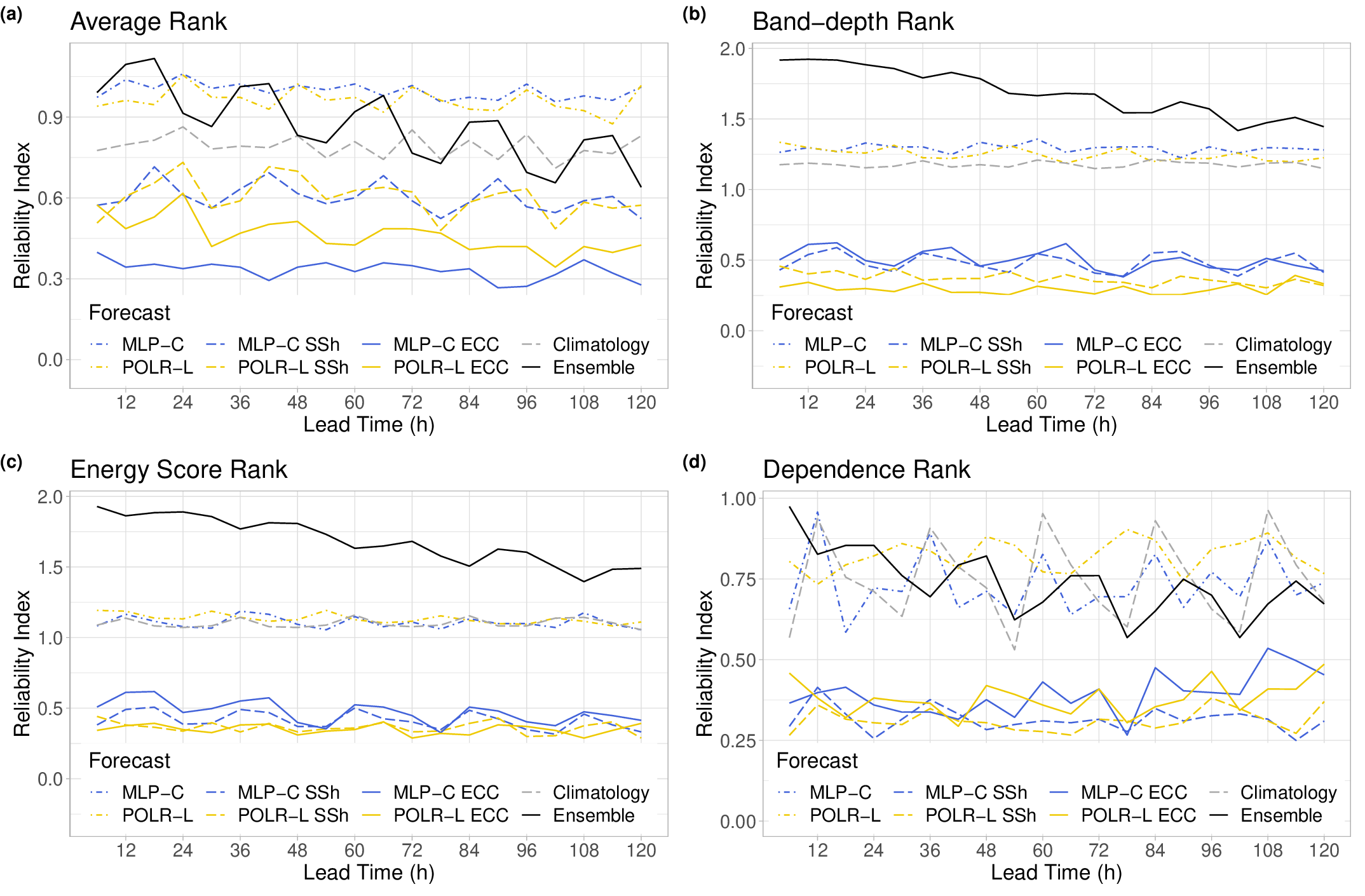, width=\textwidth}
\end{center}
\caption{Reliability indices of rank histograms of raw, climatological and post-processed multivariate visibility forecasts based on CAMS extended MLP-C and POLR-L models as functions of the lead time.}
\label{fig:relind}
\end{figure}

In addition to the previous analyses, we also generated multivariate rank histograms to thoroughly assess the differences in calibration of the individual forecasts (see Figure \ref{fig:rankhist}). In general, properly calibrated forecasts result in flat histograms, and the deviation from uniformity can be quantified by the reliability index \eqref{eq:relind}. In principle, one can have an instant ranking of the various multivariate predictions with the help of the reliability indices, displayed in Figure \ref{fig:relind} (the smaller the better); however, the interpretation of the different rank histograms can vary depending on the functions applied to determine the pre-ranks (see Section \ref{subs3.4}), as these functions may be more sensitive to detecting different deficiencies. Therefore, as argued by \cite{tsh16} and \cite{azg24}, using multiple pre-rank functions to evaluate the performance of the forecasts could be beneficial, providing a more comprehensive comparison. In the case of naive MLP-C and POLR-L models, hump-shaped average- and band-depth rank histograms indicate overestimated correlations, which is also supported by the skewed energy score- and dependence rank histograms. To a smaller extent, the same applies to the multivariate climatology. Multivariate calibration results in a substantial improvement, especially in the case of MLP-C ECC and POLR-L ECC forecasts. According to Figure \ref{fig:relind}, in terms of the average ranks, these two methods provide by far the lowest reliability indices and are very competitive with the MLP-C SSh and POLR-L SSh approaches in the three other cases, as well.

\section{Conclusions}
\label{sec5}
The current work is a direct continuation of \citet{bl23} where location-specific (univariate) discrete post-processing of visibility was studied. To get a deeper insight into the behaviour of this complex and hardly predictable weather quantity, two main questions are investigated. First, we assess whether the state-of-the-art multivariate post-processing approaches which were approved to be successful in the case of more traditional variables such as temperature, wind speed, or precipitation \citep[see e.g.][]{llhb23}, can reliably restore spatial dependencies that lost during separate calibration of visibility ensemble forecasts for each SYNOP station. Furthermore, we also investigate whether the use of CAMS visibility predictions as additional covariates in post-processing models can significantly improve the forecast skill.

In the case of the best-performing univariate post-processing approaches (MLP-C and POLR-L) from the methods investigated by \citet{bl23}, the addition of CAMS predictions to the input features results in a significant 1 -- 5.25\,\% improvement in mean CRPS and  0.27 -- 1.37\,\% in mean LogS. Note, that for both scores the gain displays a decreasing trend as the lead time increases.

Joint multivariate post-processing of forecasts for all 30 investigated SYNOP observation stations is performed with the help of the two-step ECC and SSh approaches utilizing the dependence structure of the raw vector ensemble forecasts and historical observations, respectively. In fact, in terms of ES and VS, there are no visible differences in skill between the two dependence template selection methods which substantially outperform the reference independently calibrated (naive) multivariate forecasts. Compared to forecasts based only on the ECMWF predictions, utilizing CAMS results in an additional 0.58 -- 4.11\,\% improvement in energy score for all investigated models. The VS focusing on the correct specification of correlations shows a different picture. In the case of the independent reference methods, the advantage of CAMS extended forecasts is around 5\,\%; however, the corresponding MLP-C forecasts underperform climatology for almost all lead times. In contrast, for the ECC and SSh methods, the VSS of CAMS extended predictions with respect to the models based only on the ECMWF ensemble is below 2.5\,\%; nevertheless, they outperform climatology by more than 20\,\%. From the competing multivariate predictions, the ECC-corrected MLP-C and POLR-L methods display the best overall performance.

Our case study demonstrated the usefulness of CAMS visibility predictions as additional covariates in the discrete post-processing of visibility ensemble forecasts. Furthermore, it showed the efficiency of the simplest two-step multivariate post-processing methods in capturing spatial dependencies in the case of this particular weather quantity.

A natural direction of further studies is the utilization of CAMS forecasts and/or other visibility-related covariates in post-processing methods treating visibility as a continuous quantity. Possible parametric candidates are the BMA approach of \citet{cr11}, where the predictive distribution for visibility is a mixture of beta laws, and the more recent censored gamma and censored truncated normal mixture model proposed by \citet{bb24}. However, the most straightforward extension of the present work would be the adaptation of the classification and interpolation-based approach of \citet{sswh20} to visibility forecasts.

\bigskip
\noindent
{\bf Acknowledgments.} \ 
The authors gratefully acknowledge the support of the \'UNKP-23-3  New National Excellence Program of the Ministry for Culture and Innovation from the source of the National Research, Development and Innovation Fund and of the  National Research, Development and Innovation Office under Grant No. K142849. They are also indebted to Martin Leutbecher and Zied Ben Bouall\`egue for the inspiring discussions and for providing the ECMWF and CAMS visibility data.

\end{document}